# AI Ethics in Smart Healthcare

Sudeep Pasricha
Colorado State University

***Abstract*—This article reviews the landscape of ethical challenges of integrating artificial intelligence (AI) into smart healthcare products, including medical electronic devices. Differences between traditional ethics in the medical domain and emerging ethical challenges with AI-driven healthcare are presented, particularly as they relate to transparency, bias, privacy, safety, responsibility, justice, and autonomy. Open challenges and recommendations are outlined to enable the integration of ethical principles into the design, validation, clinical trials, deployment, monitoring, repair, and retirement of AI-based smart healthcare products.**

■ **Healthcare systems** in countries around the globe are struggling to cope with health emergencies such as the COVID-19 pandemic, provide universal health coverage, and improve general health and well-being. According to the World Health Organization (WHO), access to health services without undue financial burdens for billions of individuals by 2030 urgently requires new approaches [1]. By replicating or even replacing human cognition and learning with automated machine processes, Artificial Intelligence (AI) is widely acknowledged as one of the most promising means to achieve improved healthcare access. AI solutions are competitive with, and in some cases have surpassed, human performance in computer vision [2], natural language processing [3], and recommendation engines [4], and it is undisputed that they will revolutionize precision medicine.

Contrary to popular belief, AI in healthcare is not a recent development. In the 1970s, rule-based AI algorithms were introduced for electrocardiogram interpretation, which remain commonplace today [5]. Recent innovations in AI related to deep machine learning (ML) algorithms for pattern recognition, classification, and prediction from data have reinvigorated the use of AI in healthcare. In 2018, the US FDA (Food and Drug Administration) approved the use of an AI software platform named IDx-DR to diagnose the eye disease diabetic retinopathy [6]. In 2019, the FDA approved the AI-driven wearable medical device BrainScope TBI that helps diagnose traumatic brain injury and concussions [7]. As of 2022, more than 300 AI/ML-based healthcare devices have been approved by the FDA, and there are many more in the development pipeline [8]. Many direct to consumer AI devices for health monitoring and well-being are already available and in use by patients worldwide [9].

AI is thus actively shaping new "smart" healthcare products, with the ability to learn complex trends from large datasets, and continually improve their core algorithms with new data over the lifecycle of the product. Many innovative AI-driven technologies are beginning to emerge, such as virtual health assistants, personalized therapeutics, smart digital pills, and remote robotic surgeries on the horizon. However, AI is prone to misuse when deployed inappropriately (e.g., because of over-trust), or disuse when it should be used but is not (e.g., because of under-trust) [10]. AI can be abused when implemented without due regard for the medical practitioners' capabilities or the patients' interests [11]. AI algorithms may also be unintentionally biased in their diagnosis due to biases in their training datasets or produce outcomes that are difficult for experts to comprehend. AI in smart healthcare thus raises many new ethical concerns related to transparency, bias, responsibility, safety, privacy, justice, and autonomy, as will be discussed later.

The term ethics refers to the area of philosophy concerned with the study of what is right and wrong. The sub-field of normative ethics focuses on theories that provide general moral rules governing our behavior [12]. There are three popular categories of theories – deontology, teleology (consequentialism), axiology (virtue ethics) – that are relevant for resolving ethical dilemmas, such as those that accompany the use of AI-driven healthcare products. *Deontology* makes use of rules and processes to guide ethical decisions. These rules imply that acts must be performed regardless of their consequences, e.g., the truth should always be told, no matter what its consequence. In contrast, *consequentialism* emphasizes outcomes rather than the process. It is an outcome-based approach that defines ethical behavior based on consequences (good or bad). Ethical decisions are ones that create the most good. Utilitarianism is a popular consequentialist approach which aims for the maximum good for the most number of individuals. Lastly, *virtue ethics* is guided by the question, "If I was a good person, what would I do?" Key to this theory is a reliance on core ethical virtues, such as justice, moderation, loyalty, compassion, and courage to guide ethical actions. The focus in virtue ethics is therefore on virtuous personal character when it comes to making decisions rather than universal rules or consequences. These theories are widely reflected in different spheres of aspirational modern societies and in the codes and values of professional institutions

DIFFERENCES IN MEDICAL AND AI ETHICS

Practitioners in the medical domain are well-versed with the four principles of clinical biomedical ethics: justice (fair, equitable, and appropriate treatment of patients), respect for autonomy (relating to a patient's capacity to self-determine), beneficence (obligation to act for the benefit of the patient), and non-maleficence (avoiding harm to the patient) [13]. Principlism in medicine represents a theoretical moral framework that combines traditional ethical standards with the requirements of practitioners, research ethics committees, and medical institutions for practical ethical decision-making [14]. While this ethical principlism provides guidance for clinical decision-making and setting health policy [15], there are four important differences between medicine and AI development that suggest that a principled approach for AI development (for healthcare and other contexts) may have limited impact on its design and governance. Compared to medicine, AI development lacks: *(1) Common goals and fiduciary duties:* While medicine is broadly guided by the common goal of promoting the health and well-being of patients, AI development for healthcare is driven by the private technology sector where there are constant pressures to maximize profit, reduce costs, and make decisions that prioritize the interests of the company above all else. In medicine, as in other formal professions, regulatory frameworks and professional codes have established fiduciary duties towards patients, which are facilitated by common goals and values within the profession and enforced through sanctions and self-governance [16]. This enables a principled approach to ethical decision-making by requiring practitioners to promote their clients' best interests. In contrast, AI development is not a formal profession; thus equivalent fiduciary relationships and governance mechanisms do not exist for AI developers working on developing solutions across various healthcare contexts, such as AI models for detecting early signs of sepsis (a life-threatening response to infection), brain hemorrhage detection from computed tomography (CT) scans, and medical insurance decision-making; *(2) Professional history and norms:* Medicine benefits from a long professional history and norms of good behavior, with enduring standards set in the Hippocratic oath, the Declaration of Helsinki, the Belmont Report, and other accounts of professionalism in health serving as a basis for ethical clinical decision-making and medical research. While several AI developers may be ACM members which has a longstanding ACM code of ethics, the code does not provide specific actionable strategies to deal with real-world ethical dilemmas that may be encountered by AI developers. An established profession that can draw on a rich history and established ethical culture and documented norms of good practice is best placed to conceptualize and debate new ethical challenges. Given the relatively shorter history of AI development compared to other professions, its practitioners currently lack this capability; *(3) Methods to convert principles to practice:* Over time, medicine has developed ways of converting high-level principles into norms of good practice. Ethics review committees, professional boards and societies, codes of conduct, licensing and accreditation and schemes, and other mechanisms have helped determine the ethical acceptability of everyday practice by evaluating difficult cases, identifying unacceptable behavior and punishing bad actors. AI development, being a relatively recent phenomenon, does not currently have such empirically proven methods (painstakingly derived in non-AI medical contexts over many decades via iterative methodical refinements with ethical review committees, professional boards and societies, regulatory authorities, etc.) to convert principles into practice in real-world scenarios; *(4) Professional and legal accountability:* Medicine is governed by legal frameworks that support professional standards and provide patients with redress options for negligent behavior, including ethics committees, malpractice law, and professional medical boards. Being subject to regulation incentivizes medical institutions to ensure that these standards are upheld [17]. Legally supported accountability mechanisms provide a strong external stimulus for health professionals to fulfil their fiduciary duties and promote self-governance by clearly linking bad behavior and professional sanctions (e.g., losing one's license to practice), and allowing patients to make claims against negligent members of the profession [14]. AI development today is often outsourced to software professionals that may not be aware of the ways in which their models are customized and/or deployed across products, including, but not limited to, medical devices. With the exception of certain types of risks, e.g., data privacy violations governed by laws protecting data access, AI development today does not take into account the risks related to diverse deployment contexts and lacks accountability mechanisms for developers involved with AI design and customization.

All of these differences between medical and AI ethics mean that patients cannot trust AI developers to act in their best interests when developing AI solutions for smart healthcare. The next section describes some of the specific ethical challenges with AI in healthcare. The subsequent section provides recommendations on how the differences between medical and AI ethics can be bridged in the near future.

ETHICS IN AI-DRIVEN HEALTHCARE

The deployment of AI in healthcare exacerbates existing ethical concerns in medicine and raises new ones. The relevant ethical challenges include transparency, responsibility, bias, privacy, safety, autonomy, and justice. We consider these seven ethical challenges as the most pressing of the concerns created by the use of AI in healthcare. Some of these challenges also have been raised with the growing use of AI in other application areas, particularly the automotive [18] and robotics [19] domains.

*Transparency:* AI-based healthcare devices are often black boxes due to the proprietary nature of the AI algorithms. This raises the question of how physicians can be transparent with their patients while working with systems that are inherently not entirely transparent [20]. If physicians cannot explain how a healthcare AI device arrived at a decision for a specific patient, to what extent should they rely on these solutions? Consider the use of Watson for Oncology which was widely used in China for health diagnosis via image recognition. It was later found that the underlying AI algorithms were primarily trained on a Western dataset leading to poor results for Chinese patients compared to Western patients [21]. Transparency (and "explainable AI") is thus critical to prevent systemic misdiagnosis and other undesirable outcomes. But even developers may have a hard time explaining why their complex AI algorithms behave the way they do. When hospitals purchase these black-box AI healthcare devices for a considerable amount of money and encourage their physicians to use them, how can physicians do so in an ethical manner?

*Responsibility:* If a decision made by an AI healthcare solution leads to a negative outcome for a patient, who should be responsible and liable to prevent it from happening again? In the medical field, moral responsibility has been traditionally spread transparently across stakeholders so that the causal chain of an outcome can be easily traced for a positive outcome or prevented from repeating for a negative outcome [22]. In a healthcare system reliant on AI-driven medical devices, a single AI tool may involve many people and organizations for designing AI accelerator hardware, AI algorithms, collecting and brokering training datasets, and data pre/post-processing, making this transparent allocation of responsibility very challenging. Chains of stakeholders across diverse AI-healthcare solutions, such as software-centric AI-driven virtual nursing assistants that can be accessed across various Internet-connected platforms or AI-based robotic surgeons that combine hardware, software, and their co-design, can be complex and opaque. Today's global ecosystem of outsourced hardware fabrication, software design, system testing, and client support makes accountability and responsibility difficult to assign for many of the emerging AI-based healthcare solutions.

*Bias:* If AI algorithms are trained on biased data, they may lead to increased disparities in healthcare. As an example, a physician may miss acral lentiginous melanoma in Black patients, which is a common form of melanoma in people with darker skin, because they trust an AI algorithm for cancer screening that was largely trained on white skin [23]. Biased AI algorithms can have serious implications, as highlighted by the case of an AI algorithm used by large healthcare systems and payers to guide health decisions for almost 200 million people in the U.S. annually. The algorithm incorrectly assigned the same level of risk to Black and White patients, despite Black patients in the dataset being much sicker [24]. The racial bias was a result of the algorithm using healthcare costs instead of illness as a measure of the level of health needs. As Black patients' healthcare related spending was lower, the algorithm incorrectly concluded that Black patients were healthier. AI algorithms may also learn to prioritize patients they predict will have better outcomes for a particular disease, which can again have a discriminatory effect on economically less well-off communities [25]. How should AI developers minimize such bias? And how should physicians deal with bias in black box AI algorithms, reactively and proactively? These are critical ethical concerns that must be addressed.

*Privacy:* The collection of vast volumes of data from wearables and health apps creates many opportunities for compromising patient privacy. Consider the case of the Dinerstein vs. Google lawsuit brought on by Matt Dinerstein, a patient at the University of Chicago Medical Center (UCMC), individually and on behalf of other patients, against Google and UCMC [26]. In 2017, UCMC and Google established a partnership to use AI to predict medical events, such as hospital readmissions. Between 2009 and 2016, Dinerstein claimed that UCMC transferred hundreds of thousands of medical records to Google, without obtaining patients' express consent. However, a federal judge dismissed this lawsuit in 2020 on the grounds that Dinerstein had failed to demonstrate damages. This case highlights the insufficient protection of health data privacy and the challenges of preserving privacy with AI-based healthcare products that are backed by tech giants and institutions that can bankroll powerful legal teams.

*Safety:* AI algorithms in healthcare products must not cause harm. Today, AI apps and assistants are designed to limit unnecessary doctor visits. If a patient with chest pain is diagnosed by the AI algorithm as having gastroesophageal reflux disease (acid reflux) and provided a recommendation for an antacid, when in fact they are suffering from a heart attack, this is unacceptable. Conversely, treating most instances of chest pain as heart attacks will lead to false positives that unduly burden the healthcare system and increase costs for patients. The FDA has begun to regulate software-based healthcare products and made efforts to develop a software precertification program to ensure their safety and effectiveness [27]. However, regulating the use of AI algorithms that continue to learn and adapt over time remains unaddressed. Moreover, many AI algorithms currently in use are not subject to FDA review, such as certain AI-driven clinical decision support software and many healthcare apps and chatbots. Often, AI-based medical devices are trained to minimize mathematical "loss functions", and not explicitly and by default to minimize harm to patients. Thus, safety with AI in healthcare remains an open challenge.

*Autonomy:* The use of AI creates new issues with informed consent. If a physician uses AI-based healthcare products to determine a treatment plan, does a patient have the right to be informed that an AI was part of the decision-making process? If yes, what should the patient be told? Does the patient need to be informed about the dataset used to train the AI, and whether it was based on real health record data or synthetic data? It is important to determine what informed consent with AI should look like for a patient, and how it would extend general treatment consent that patients sign today in medical contexts.

*Justice:* The just allocation of healthcare resources is an ongoing challenge which AI algorithms will complicate. Consider the case of Tammy Dobbs, a patient with cerebral palsy who was allocated 56 hours of care per week in Arkansas as part of a state program. In 2016, the state began to rely on an AI algorithm to allocate caregiving and her care was reduced to 32 hours, which was insufficient according to her [28]. As the decisions made by AI algorithms are often invisible, it will complicate decisions related to just allocation of health resources. Many of the most vulnerable patients may not have access to AI-based healthcare devices and tools for diagnosis and treatment. Then the question is not only how to justly allocate healthcare but also how AI can be designed to promote justice rather than to subvert it.

ETHICAL AI-BASED SMART HEALTHCARE

Given the many urgent ethical challenges that AI-based smart healthcare brings, as discussed in the previous section, there is a need for approaches that encompass all the stakeholders in the impacted ecosystem. Multiple strategies will need to be employed to address the ethical challenges of transparency, responsibility, bias, privacy, safety, autonomy, and justice with AI-based healthcare. Each of the strategies discussed below has the promise to overcome multiple ethical challenges.

**Ethical AI education**

Many universities and research institutions are beginning to emphasize topics related to ethics in their technical curricula with the goal of raising ethical awareness in developers, programmers, and engineers of AI and other technologies [29]. Many technology companies working with AI are implementing training modules on ethics for their employees, e.g., ethical foresight analysis, to educate designers and managers in predicting potential ethical issues and the consequences of specific technologies [30]. These modules must include case studies at the intersection of AI and medicine, with an emphasis on transparency, bias, safety, privacy, responsibility, autonomy, and justice. The case studies should highlight the appropriateness of relying on specific normative ethics theories such as deontology, teleology, or axiology, based on the situation, as part of applied ethics analyses, to resolve ethical dilemmas. Such efforts must also be extended to the curricula of medical students [31] who will be using and recommending the AI products, so they can comprehend the ethical harms of AI misuse, disuse, and abuse, as part of a careful risk-benefit analysis.

**Ethicist in the AI development loop**

AI development today often occurs in competitive environments that value speed and efficiency and, in commercial settings, also profit. Ethical considerations can be ignored if they conflict directly with commercial incentives [32]. It is therefore imperative to have one or more ethicists that are part of the development process, who are tasked with identifying and integrating ethical capabilities into AI-based healthcare products. As the development team may not be able to perceive every ethical issue, regular interactions between the ethicists and technically focused engineers would reduce the risk of ethical issues and conflicts being overlooked. Without being overly prescriptive, ethicists could help explain and clarify complex ethical issues to enable a clearer understanding of them and use methods of ethical reasoning to challenge or justify a position or course of action. To enable more transparency and mitigate bias with AI algorithms, the ethicists could advocate for important decisions to be made as a 'white box' rather than a 'black box' so that stakeholders can scrutinize and understand how the algorithm makes decisions and allow for social accountability. Such a collaborative ethics approach has been successfully used in the genomics field [33] and could greatly benefit the ethical design of emerging AI-based healthcare products.

**Ethical healthcare product lifecycle analysis**

Ethical challenges can arise at many instances over the entire lifecycle of an AI-based healthcare product. This analysis should include identification of, engagement with, and explicit communication about the diverse values and perspectives of all stakeholders – such as developers, sales reps, technicians for installation, physicians, hospitals, repair/debug specialists, and patients – while supporting systematic and thorough reflection and reasoning about the ethical issues. The reflection on ethical issues should go beyond teleological impacts and consequences of the healthcare product and include considerations at all stages of the product lifecycle, including the earliest stages of initial conceptual design and market analysis (to determine the ethics of the multiple pathways to innovation), design, validation, clinical trials, deployment, lifecycle monitoring, repair, and retirement. Developments in ethical product lifecycle analysis from the agricultural biotechnology field [34] are particularly relevant for AI-based healthcare products. An example of this is the Ethical Matrix method, which is a tool to evaluate the intersection of three normative ethical principles (respect for well-being, autonomy, and justice) with four relevant stakeholder groups (the treated organisms, producers, consumers, and environment). The applicability to AI-based healthcare product development could be imagined with application of the same important principles to the relevant stakeholder groups of patients, smart healthcare companies, surgeons, and hospitals, and concerns that encompass economic, regulatory, sustainability, and societal factors.

**Ethical AI-based smart healthcare policy**

Translating ethical principles into practice for AI-driven healthcare products will require regulatory support from the government, the medical profession, and industry. Legally binding and highly visible accountability structures must be established, along with professional and institutional norms that define the key requirements for inclusive design, documentation of datasets and AI models, transparent ethical review, and independent ethics-based auditing [35]. Current regulatory frameworks used by the US FDA involve reviewing medical devices through a premarket pathway, such as premarket clearance (510(k)), De Novo classification, or premarket approval. The FDA may also review and clear modifications to medical devices, including software as a medical device, depending on the significance or risk posed to patients of that modification. The US FDA has however acknowledged [8] that "The FDA's traditional paradigm of medical device regulation was not designed for adaptive artificial intelligence and machine learning technologies. Under the FDA's current approach to software modifications, the FDA anticipates that many of these artificial intelligence and machine learning-driven software changes to a device may need a premarket review." The US FDA published an action plan in Jan 2021 [8] for using AI in healthcare devices, expressing the intent to establish "good machine learning practices," oversight of how algorithms behave in real-world scenarios and development of research methods for rooting out bias. However, these intentions have not yet been integrated into a formal regulatory framework. More recently, in Jun 2022, the FDA published guidelines on AI in radiological devices, requiring companies to outline how the technology is supposed to perform and provide evidence that it works as intended. In Sep 2021, the European Commission (EU) also published a proposal for a legal framework as part of the AI Act, to promote the development of an ecosystem of trust with AI [36]. The legislation presented a risk-based regulatory approach to AI, identifying healthcare as a high-impact sector for AI and requiring AI-based healthcare devices to fulfill the requirements of the AI Act as well as those already set under the existing EU Medical Device Regulation. These are steps in the right direction. However, many gaps exist across these regulatory frameworks [37]. For instance, the US guidelines do not stipulate rigorous prerequisites and manufacturer responsibility like the EU regulations do, for lifecycle regulation and algorithmic bias mitigation in AI-based healthcare devices. Conversely, the US guidelines call for promoting greater transparency and publishing summaries for each approved healthcare device, which is not the case in EU where their database on medical devices (Eudamed2) is not publicly accessible.

To inspire long-term recognition of ethical principles, formal recognition of AI development for healthcare as a profession with equal standing as for other high-risk professions may be necessary. It is a regulatory anomaly that professions that provide a public service are licensed, but the profession responsible for designing technical systems to augment and replace human decision-making in public healthcare remains unlicensed [32]. Lastly, many AI ethics initiatives to date have created professional codes of ethics that address design goals and the behaviors and values of individuals but ignore the validity of specific applications and their underlying organizational and business interests [38]. This approach conveniently steers debate towards the wrongdoings of unethical individuals, and away from the overall failure of business models and unethical organizations. Regulations will be essential to hold erring businesses accountable. Such regulations can also target other big issues such as fair distribution of healthcare benefits, protecting equality of care, and promoting ethical societal values in the era of AI-driven healthcare.

## CONCLUSIONS

In this article, we reviewed the landscape of ethical challenges facing AI integration into healthcare products. AI introduces many new ethical concerns for healthcare products related to transparency, bias, responsibility, safety, privacy, justice, and autonomy that go beyond traditional medical ethics principles, and must be addressed during the design, deployment, and use of these systems. Broadly speaking, healthcare systems in many ways act as the core of modern societies. If ethical mistakes are made in these early days of adoption and implementation of AI in healthcare, the fallout could undermine public trust, which could be devastating for the healthcare industry and also cause patients to look for their healthcare from outside of formal systems where they may encounter significant risks. To address this challenge, we discussed multiple approaches that will be crucial to promote the ethical integration of AI in healthcare.

## ACKNOWLEDGEMENTS

This work was supported by National Science Foundation (NSF), through grant CNS-2132385.

## ■ REFERENCES

**Sudeep Pasricha** received his Ph.D. in Computer Science from UC Irvine in 2008. He is a Professor in the Department of Electrical and Computer Engineering and Director of the Embedded, High Performance, and Intelligent Computing (EPIC) Lab at Colorado State University. His research interests relate to embedded and IoT systems, with an emphasis on designing innovative hardware architectures, software algorithms, and hardware-software co-design techniques for fault-tolerant, energy-efficient, secure, real-time, and ethical computing. He is a Senior Member of IEEE and Distinguished Member of ACM. Contact him at sudeep@colostate.edu.